\documentclass[aps,prc,twocolumn,superscriptaddress]{revtex4} 
\usepackage{graphicx}
\usepackage[varg]{txfonts} 
\usepackage[latin1]{inputenc}

\newcommand{\oeq}{\begin{equation}}
\newcommand{\ceq}{\end{equation}}
\newcommand{\oeqn}{\begin{eqnarray}}
\newcommand{\ceqn}{\end{eqnarray}}

\newcommand{\hb}{\hbar}

\begin{document}
\title{Effects of Nuclear Structure on Quasi-fission}

\author{C. Simenel}\email[]{cedric.simenel@anu.edu.au}
\affiliation{CEA, Centre de Saclay, IRFU/Service de Physique Nucl\'eaire, F-91191 Gif-sur-Yvette, France}
\affiliation{Department of Nuclear Physics, Research School of Physics and Engineering, 
  Australian National University, Canberra, ACT 0200, Australia}

\author{A. Wakhle}
\affiliation{Department of Nuclear Physics, Research School of Physics and Engineering, 
  Australian National University, Canberra, ACT 0200, Australia}

\author{B. Avez}
\affiliation{Universit\'e Bordeaux 1, CNRS/IN2P3, Centre d'\'Etudes Nucl\'eaires de Bordeaux Gradignan, 
CENBG, Chemin du Solarium, BP120, 33175 Gradignan, France}
\affiliation{CEA, Centre de Saclay, IRFU/Service de Physique Nucl\'eaire, F-91191 Gif-sur-Yvette, France}

\author{D. J. Hinde}
\affiliation{Department of Nuclear Physics, Research School of Physics and Engineering, 
  Australian National University, Canberra, ACT 0200, Australia}

\author{R. du Rietz}
\affiliation{Department of Nuclear Physics, Research School of Physics and Engineering, 
  Australian National University, Canberra, ACT 0200, Australia}

\author{M. Dasgupta}
\affiliation{Department of Nuclear Physics, Research School of Physics and Engineering, 
  Australian National University, Canberra, ACT 0200, Australia}

\author{M. Evers}
\affiliation{Department of Nuclear Physics, Research School of Physics and Engineering, 
  Australian National University, Canberra, ACT 0200, Australia}

\author{C. J. Lin}
\affiliation{Department of Nuclear Physics, Research School of Physics and Engineering, 
  Australian National University, Canberra, ACT 0200, Australia}

\author{D. H. Luong}
\affiliation{Department of Nuclear Physics, Research School of Physics and Engineering, 
  Australian National University, Canberra, ACT 0200, Australia}

\begin{abstract}
The quasi-fission mechanism hinders fusion of heavy systems because of a mass flow between the reactants, leading to a re-separation of more symmetric fragments in the exit channel. A good understanding of the competition between fusion and quasi-fission mechanisms is expected to be of great help to optimize the formation and study of heavy and superheavy nuclei. 
Quantum microscopic models, such as the time-dependent Hartree-Fock approach, allow for a treatment of all degrees of freedom associated to the dynamics of each nucleon. This provides a description of the complex reaction mechanisms, such as quasi-fission, with no parameter adjusted on reaction mechanisms.
In particular, the role of the deformation and orientation of a heavy target, as well as the entrance channel magicity and isospin are investigated with theoretical and experimental approaches. 
\end{abstract}
\maketitle
\section{Introduction}

The formation of the heaviest nuclei usually involves fusion-evaporation reactions~\cite{hof00,mor07,oga06,hof07}. 
The latter are strongly hindered in the case of heavy ion reactions by two 
competing mechanisms: $(i)$ the quasi-fission (QF) process, and $(ii)$ the statistical fission of the compound nucleus (CN). 

Quasi-fission occurs in the early stage of the collision~\cite{boc82,tok85,she87}, when the two reactants form a di-nuclear system, that is, two fragments linked by a neck. An important nucleon transfer usually occurs from the heavy fragment toward the light one. The two fragments then re-separate with more mass symmetry than the entrance channel, without forming a compound nucleus. 

Typical QF times are shorter than $10^{-20}$s~\cite{boc82,tok85,she87,rie11}.
These times have to be compared with fusion-fission times which can be longer than $10^{-16}$~s~\cite{and07}.
This shows that the two mechanisms are of very different nature. 
In fact, the QF process is a  dynamical mechanism depending on the characteristics of the entrance channel, while CN fission is pure-ly statistical and is determined by  temperature and angular momentum only.
 
The QF mechanism is responsible for the fusion hindrance observed in heavy systems~\cite{gag84}. In these reactions, an additional energy above the Coulomb barrier, sometimes called ''extra-push'' energy~\cite{swi82}, is needed for the system to fuse and form a CN. 
Note that lighter systems may also exhibit quasi-fission, although with a smaller probability. 
For instance, quasi-fission has been observed in $^{16}$O,$^{32}$S+$^{238}$U \cite{hin96,itk11,nas07}, and in $^{32}$S+$^{208}$Pb~\cite{nas07}. 

The QF process is known to be affected by nuclear deformation and orientation at energies close to the fusion barrier~\cite{hin95,liu95,hin96,oga04,kny07,hin08,nis08}. 
The role of isospin has also been investigated theoretically~\cite{kal11}.
Recently, the influence of entrance-channel magicity and isospin on quasi-fission has been investigated~\cite{sim12a}. Magic shells are indeed expected to generate  ''cold valleys'' in the potential energy surface, favouring the formation of a compact CN~\cite{san76,faz05,ari06}. 
In addition, magic nuclei are difficult to excite, reducing energy dissipation, and, then, allowing more compact di-nuclear systems~\cite{hin05,arm00}.

Many experimental data on QF are now available for comparison with theoretical models in order to test their predictive power. 
This is indeed crucial to have reliable theoretical models in order to drive future experiments on heavy elements formation. 
Macroscopic approaches have been thoroughly used  in the past~\cite{zag01,ada03}. 
In addition, the recent increase of computational power allowed micros-copic descriptions of nuclear dynamics.  For instance, systems as heavy as actinide collisions have been studied with microscopic approaches~\cite{tia08,gol09,zha09,ked10,sim11b}.

In particular, the time-dependent Hartree-Fock (TDHF) theory~\cite{dir30} is particularly well suited at low energies where a proper treatment of the interplay between reaction mechanisms and nuclear structure is crucial.
Indeed,  TDHF calculations treat both dynamics and ground-state structure on the same footing, i.e., with the same energy density functional (EDF) as the only phenomenological input. 
As a result, the TDHF approach has been successful in describing several reaction mechanisms, such as fusion, nucleon transfer, and deep-inelastic collisions (see Ref.~\cite{sim12b} for a review). 

We first illustrate the fusion hindrance with TDHF calculations of fusion thresholds in heavy systems.  
Then we present an experimental study of the role of magicity and isospin on the quasi-fission mechanism.
Finally, we discuss results of a recent study of quasi-fission with TDHF calculations.

\section{Fusion hindrance in heavy systems\label{sec:hindrance}}

Recent TDHF calculations have been performed to investigate collisions of heavy systems leading to quasi-fission and formation of super-heavy compound nuclei~\cite{gol09,obe10,uma10b,ked10,sim11b,sim12c,guo12}.
In particular, fusion hindrance in heavy systems was predicted~\cite{sim12c,guo12}. 
The additional energy needed for fusion to occur has been found to be of the same order of magnitude as the extra-push energy~\cite{sim12c} determined with the phenomenological approach of Swiatecki~\cite{swi82}. 

\begin{figure}
\includegraphics[width=8cm]{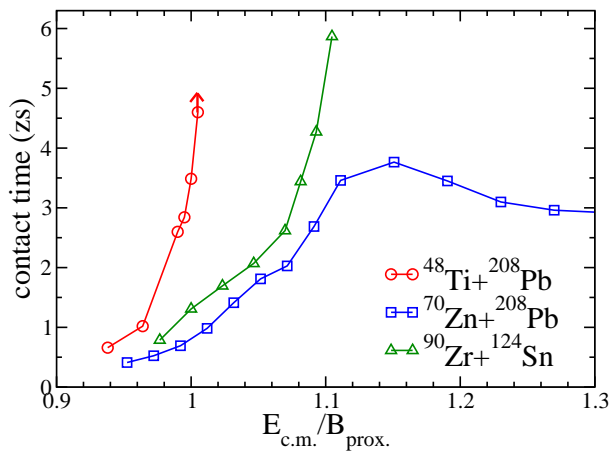} 
\caption{Contact time for  heavy-ion central collisions as a function of  center of mass energy normalised to the proximity barrier~\cite{blo77}. The arrow indicates a lower limit. 
}
\label{fig:contact}
\end{figure}

Fig.~\ref{fig:contact} shows the evolution of contact time as a function of energy for different central collisions obtained with the \textsc{tdhf3d} code~\cite{kim97}. 
Here, the contact time is determined arbitrarily as the time the system spend with a distance between the centers of mass of the fragments smaller than 15~fm.
The most asymmetric reaction, $^{48}$Ti+ $^{208}$Pb with $Z_1Z_2=1804$, produces long contact times at energies above the proximity barrier~\cite{blo77} which could be assimilated to fusion. 
In this case, no (or little) extra-push energy is needed to fuse. 

Increasing the mass of the projectile, as in the $^{70}$Zn+ $^{208}$Pb reaction with $Z_1Z_2=2460$, changes drastically this behaviour. 
In this case, we observe a slow increase of the contact time above the barrier, with a maximum of $\sim4$~zs at $\sim1.15B_{prox.}$. 
These calculations do not show any fusion for this system.
Instead, the di-nuclear system always encounters quasi-fission. 

The $^{90}$Zr+$^{124}$Sn reaction is more symmetric, but with an intermediate charge product $Z_1Z_2=2000$. 
At and up to $10\%$ above the barrier, the collision time increases slowly, similarly to the $^{70}$Zn+$^{208}$Pb case, indicating that quasi-fis-sion occurs within this energy range. 
At higher energies, the TDHF calculations predict long contact times (greater than $20$~zs) which may be associated to fusion reactions. 
As a result, an extra-push  $E_X^{TDHF}\sim 23$~MeV is predicted for this system, which is of the same order of magnitude, but slightly larger, than the value given by the extra-push model $E_X^{Swi.}\sim 15$~MeV \cite{swi82}. 
Similar agreements have been obtained for other systems in Ref.~\cite{sim12c}.

Detailed investigations on the quasi-fission mechanism are mandatory to understand fusion hindrance in heavy systems. 
In particular, a good understanding of how these mechanisms are affected by entrance channel properties is of utmost importance to optimise  the formation of super-heavy elements. 

\section{Quasi-fission: interplay between shells and isospin}

\begin{figure}
\includegraphics[width=8cm]{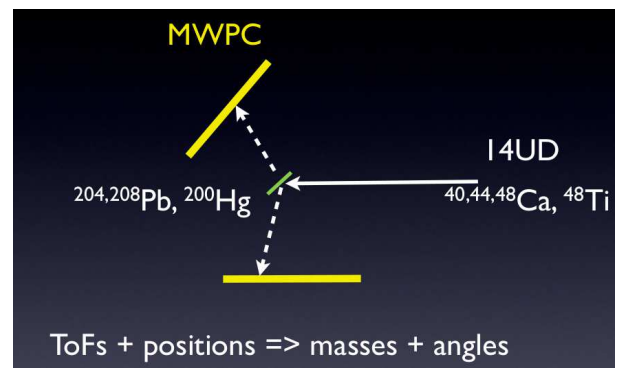}
\caption{Experimental setup for the measurement of MAD. 
\label{fig:setup}}
\end{figure}

A series of experiments to investigate the quasi-fission process have been performed recently at the Australian National University in Canberra~\cite{hin11,rie11,lin12,hin12,sim12a}.
In Ref.~\cite{sim12a}, we showed that shell effects in the entrance channel may  affect the quasi-fission process. 
This was done by measuring mass-angle distributions (MAD) of the fragments in several reactions at sub-barrier energies. 
The experimental setup is shown in Fig.~\ref{fig:setup}.
The beams were produced by the 14UD electrostatic accelerator.
Two Multi-Wire Proportional Chambers (MWPC) were used to measure time and positions of both fission fragments in coincidence (see Ref.~\cite{sim12a} for more details on the geometry of the setup). 
Time of flight (ToF) and positions were converted into fragment masses and angles using two-body kinematics.

\begin{figure*}[t]
\begin{center}
\includegraphics[width=17cm]{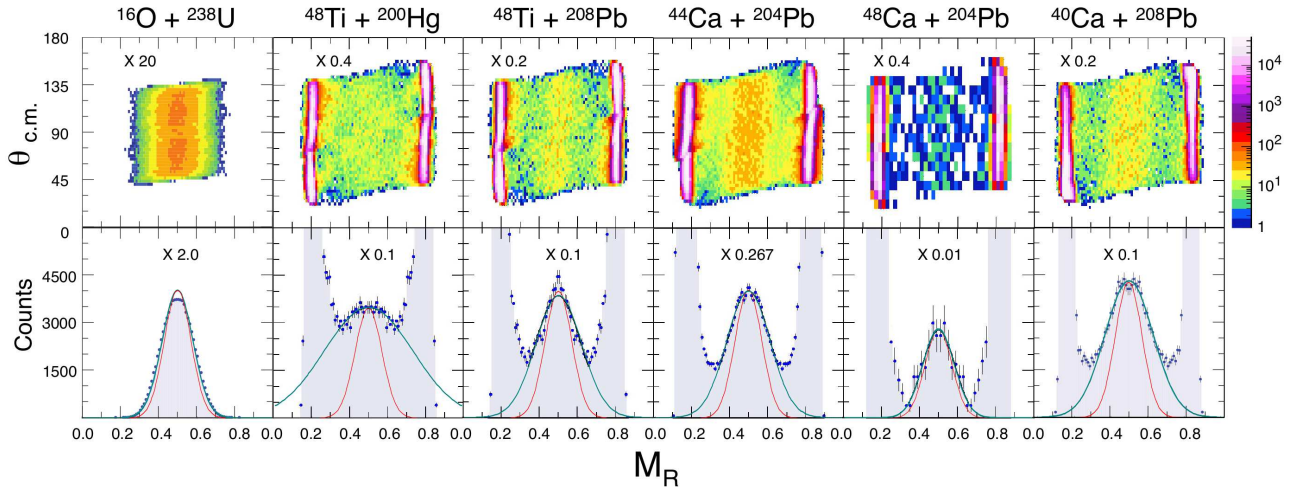}
\caption{(upper panels) Measured MAD. (lower panels)  Projected mass ratio spectra.
Gaussian fits to the region around $M_R$=0.5 are shown (turquoise lines). Gaussian functions with $\sigma_{M_R}=0.07$ (thin red lines) are shown for reference.}
\label{fig:MAD}
\end{center}
\end{figure*}

\begin{figure*}
\begin{center}
\includegraphics[width=15cm]{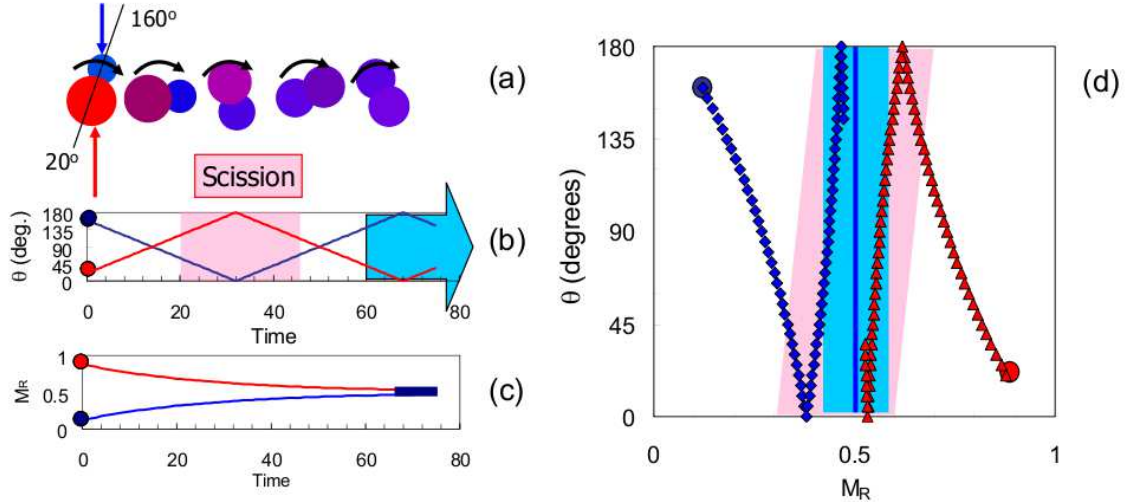}
\caption{Qualitative illustration of the distribution of quasi-fission fragments in the MAD.
(a) Non-central collisions induce a rotation of the di-nuclear system. During the rotation, nucleon transfer occurs toward symmetry. (b) Evolution of the angle as a function of time. The two colored area corresponds to two different scission times. (c) Time evolution of the fragment masses toward symmetry. (d) ''trajectory'' of the fragments in the MAD. Short scission times (pink area) lead to an angle in the MAD, whereas for longer times (blue area) such correlations disappear. Adapted from Ref.~\cite{hin11}.}
\label{fig:sketch}
\end{center}
\end{figure*}

Fig.~\ref{fig:MAD} shows the resulting MAD (up) and the projections on the mass-ratio axis (bottom). 
The mass ratio is defined as $M_R=m_2/(m_1+m_2)$ where $m_2$ (resp. $m_1$) is the mass of the fragment in the back (front) detector. 
Fission and quasi-fission fragments are located between the bands at extreme $M_R$ corresponding to (quasi-)elastic and deep-inelastic events. 
Fig.~\ref{fig:sketch} sketches the ''trajectory'' of the fragments in the MAD.
In particular, short scission times induce correlations, i.e. the distribution forms a finite angle with the $M_R=0.5$ axis. 
Note that this simple picture neglects possible shell effects in the exit channel which may enhance the production of magic nuclei such as $^{208}$Pb and ''delay'' the mass drift toward symmetry. 
In addition, long-time fission of super-heavy systems may occur via asymmetric channels~\cite{mor08,fre12}, i.e., fusion-fission does not necessarily populate the region of the MAD around the $M_R=0.5$ axis.

Correlations between mass and angle of the fission fragments can be seen on some spectra of Fig.~\ref{fig:MAD} (see, e.g., the case of $^{44}$Ca+$^{208}$Pb which has the highest statistics). 
Such correlations increase the width of the mass ratio distribution (lower panels).
We then use the width of the fission fragment mass distribution to quantify the amount of ''fast re-separation'' which could be associated to quasi-fission events. 
In particular, the $^{16}$O+$^{238}$U reaction, which is known to exhibit  only a small amount of quasi-fission \cite{hin96}, gives a width which can be considered as an upper limit for pure fusion-fission events. 

\begin{figure}
\begin{center}
\includegraphics[width=6.5cm]{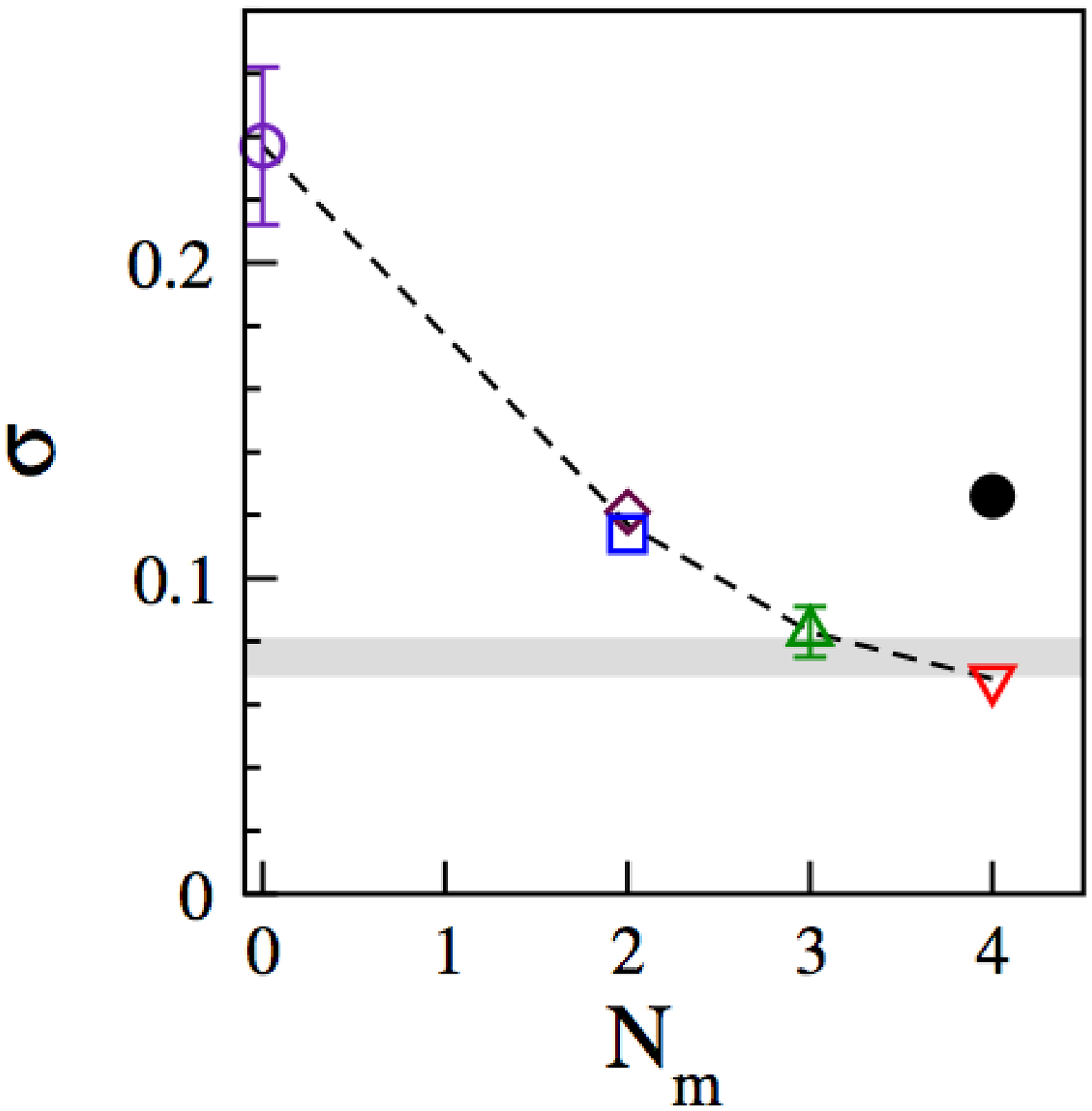}
\caption{
Widths of the mass distributions of the fission-like events as a function of the magicity (quantified by the number of magic numbers $N_m$) in the entrance channel~\cite{sim12a}.}
\label{fig:width}
\end{center}
\end{figure}

\begin{figure}
\includegraphics[width=8cm]{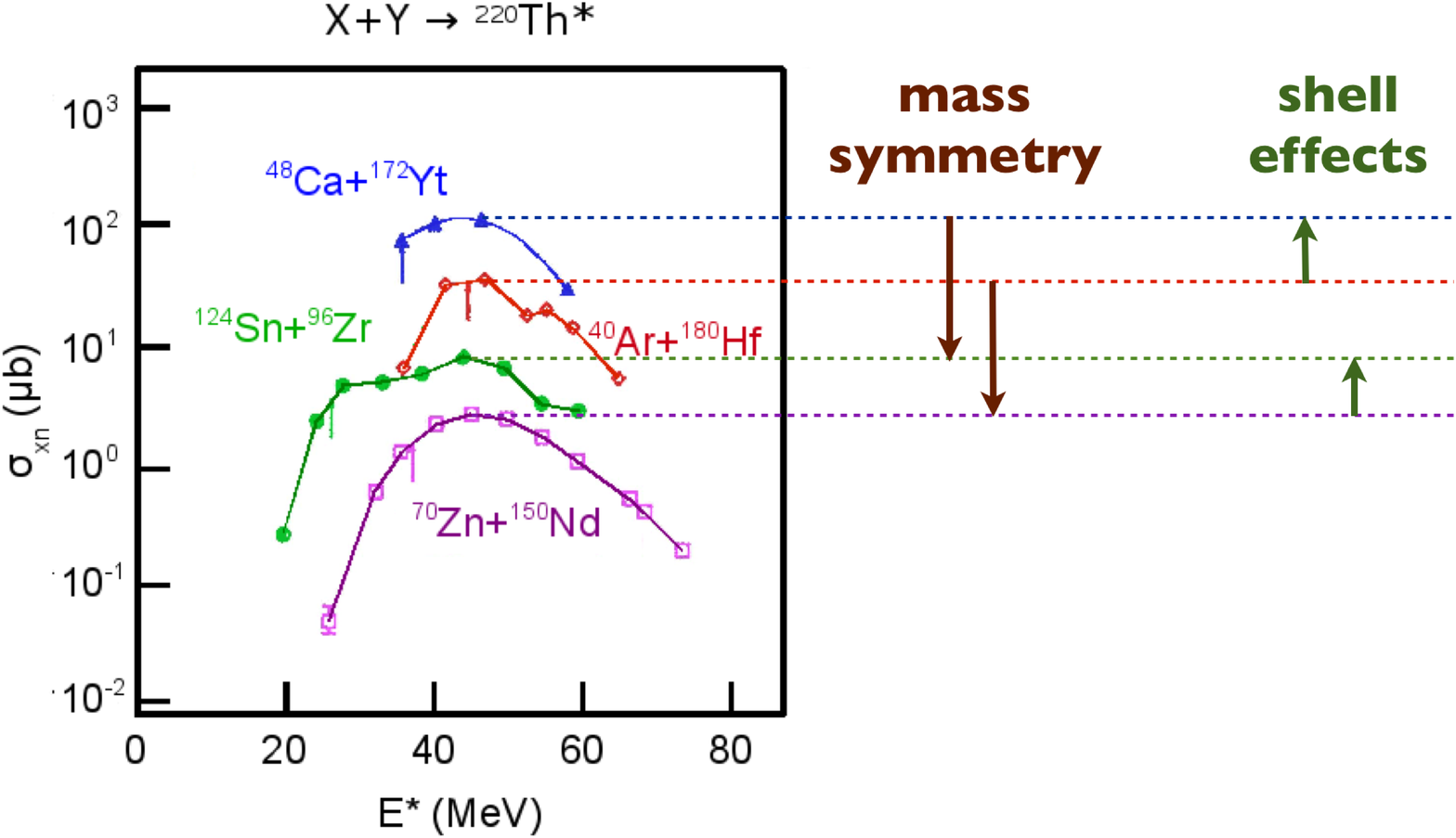}
\caption{Experimental fusion-evaportation cross-sections in reactions forming $^{220}$Th$^*$. Adapted from~\cite{sto98}.}
\label{fig:stodel}
\end{figure}

Fig.~\ref{fig:width} shows the width of the fission-like fragment mass distributions as a function of the number of magic numbers in the entrance channel. 
We  see that, apart for the $^{40}$Ca+$^{208}$Pb case which is discussed below, there is a clear link between the two quantities, i.e., the more the magicity, the smaller the width. 
This is interpreted as a hindrance of the transfer toward symmetry process due to shell-effects in the entrance channel. 

\begin{figure}
\includegraphics[width=8cm]{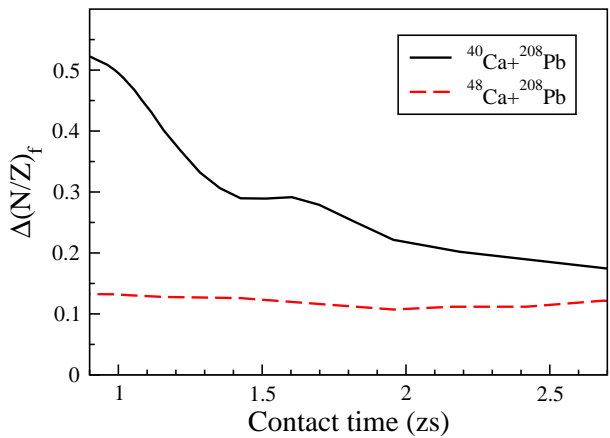}
\caption{TDHF calculations of charge equilibration: the difference between final and initial $N/Z$ of the fragments is shown as a function of the contact time. Adapted from~\cite{sim12a}.}
\label{fig:CaPb}
\end{figure}

The effect of entrance channel magicity on fusion as been investigated in Ref.~\cite{sto98}. 
Fig.~\ref{fig:stodel} shows a comparison of fusion-evaporation cross-sections for reactions forming $^{220}$Th$^*$~\cite{sto98}. 
We can see that fusion-evaporation cross-sec-tions decrease when the mass asymmetry increases. 
On the contrary, these cross-sections increase when shell effects are present in the entrance channel.
Indeed, $^{48}$Ca is doubly magic and $^{124}$Sn has a magic proton number. 
Assuming that capture cross-sections, i.e., the sum of QF and fusion cross-sections, are not sensitive to shell-effects, we conclude that shell effects in the entrance channel hinder quasi-fission while it favours fusion of the fragments.

We see in Fig.~\ref{fig:width} that the $^{40}$Ca+$^{208}$Pb does not lie on the global trend. 
In Ref.~\cite{sim12a}, we interpreted this apparent discrepancy as an effect of charge equilibration occurring in the early stage of the collision. 
This equilibration process is indeed very rapid as shown by the TDHF calculations reported in Fig.~\ref{fig:CaPb}.
The latter are performed at the experimental energy $E_{c.m.}=179.1$~MeV, and varying $L$ to obtain different collision times. 
$\Delta(N/Z)_f$ is the difference between the $N/Z$ ratio of the fragments after the collision. 
The initial value of  $\Delta(N/Z)$ is 0.54 for $^{40}$Ca+$^{208}$Pb and 0.14 for $^{48}$Ca+$^{208}$Pb.
We observe a (partial) equilibration (i.e., a reduction of $\Delta(N/Z)_f$) within $\sim2$~zs in the $^{40}$Ca+$^{208}$Pb system. 
This time is short as compared to the quasi-fission time which, according to simple simulations based on the model of Ref.~\cite{rie11}, is longer than 10~zs. 
This means that the collision partners change their $N$ and $Z$ at contact and,  as far as quasi-fission is concerned, behave like a non-magical system. 
This effect is not present in $^{48}$Ca+$^{208}$Pb which is less $N/Z$ asymmetric and does not encounter a charge equilibration (see the red dashed line in Fig.~\ref{fig:CaPb}). 
Fusion is then more probable in this reaction which preserves its magic nature in the di-nuclear system.  

We conclude that shell effects in the entrance channel hinder quasi-fission (and then, favour fusion) only for systems with small $N/Z$ asymmetry.  

\section{Quasi-fission within the TDHF approach\label{sec:QFTDHF}}

The previous studies showed the importance of the quasi-fission process as a mechanism in competition with fusion, hindering the formation of heavy systems. 
A quantum and microscopic theoretical framework able to describe properly the quasi-fission properties would be of great importance to get a deep insight into the interplay between structure properties and the QF mechanism.

We saw in Section~\ref{sec:hindrance} that quasi-fission was responsible for fusion hindrance in heavy systems. 
We now use the TDHF approach to investigate the  quasi-fission process.
Fig.~\ref{fig:contact} indicates that fusion hindrance appears in systems with larger charge products than in the $^{48}$Ti+$^{208}$Pb system (see also Ref.~\cite{sim12c}).  
In the following we study the $^{40}$Ca+$^{238}$U reaction ($Z_1Z_2=1840$) around the barrier which, indeed, leads to quasi-fission at the mean-field level~\cite{wak12,sim12c}.

\begin{figure}
\begin{center}
\includegraphics[width=8cm]{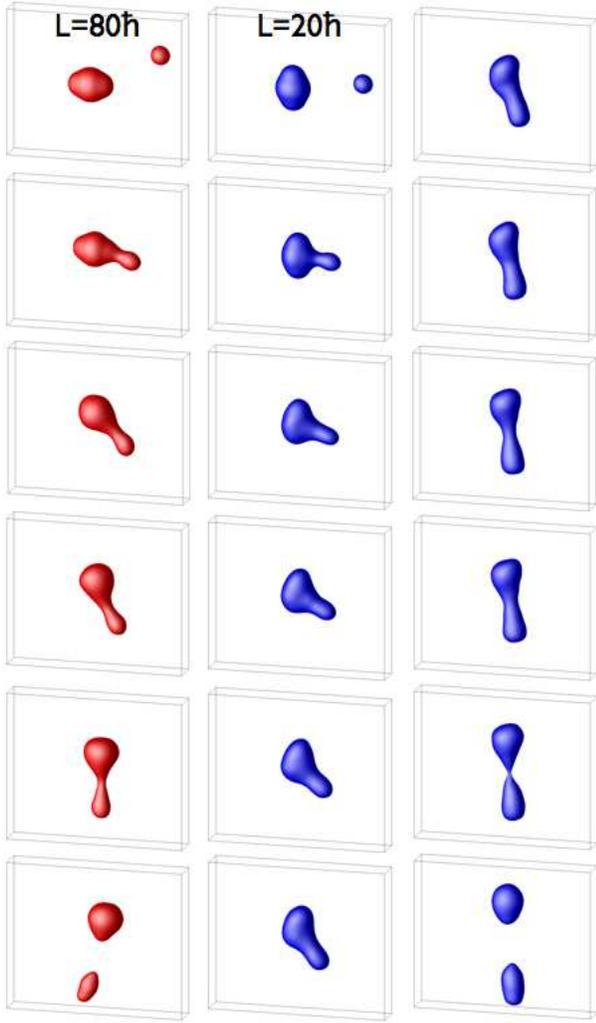}
\caption{Snapshots of the TDHF isodensity at half the saturation density ($\rho_0/2=0.08$~fm$^{-3}$) in the $^{40}$Ca+$^{238}$U system at $E_{c.m.}=206$~MeV. (left) Collision with a tip of $^{238}$U at $L=80\hb$. (middle and right) Collision with the side of $^{238}$U at $L=20\hb$. Snapshots are shown every 1.5~zs.}
\label{fig:densQF}
\end{center}
\end{figure}

Examples of density evolutions obtained with the \textsc{tdhf3d} code are shown in Fig.~\ref{fig:densQF} for this system at $E_{c.m.}=206$~MeV. 
The left column shows the density for a collision with a tip of the $^{238}$U nucleus at $L=80\hb$, leading to an average exit channel $^{55}$Mn+$^{218}$Fr. 
The middle and right columns show a collision with the side of $^{238}$U at $L=20\hb$. In this case the  exit channel is more symmetric: $^{111}$Rh+$^{167}$Ho in average. 
These reactions correspond to quasi-fission, i.e., an important multi-nucleon transfer from the heavy fragment toward the light one with several zs life-time of the di-nuclear system which is typical for QF~\cite{boc82,tok85,she87,rie11}. 
We also see that the mass equilibration (i.e., the formation of two fragments with symmetric masses) is not complete and may depend on the initial conditions.

\begin{figure}
\begin{center}
\includegraphics[width=8cm]{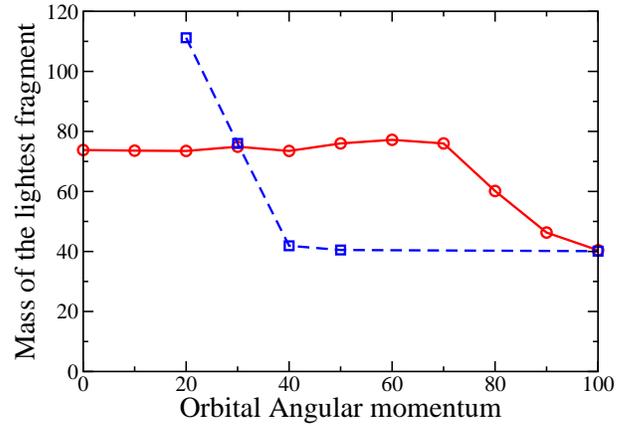}
\caption{Mass of the light final fragment in $^{40}$Ca+$^{238}$U collisions at $E_{c.m.}=206$~MeV as a function of the angular momentum. Collisions with the tip (red circles, solid line) and with the side (blue squares, dashed line) of $^{238}$U are considered.}
\label{fig:M_L}
\end{center}
\end{figure}

To get a better insight into the mass transfer mechanism, we plot in Fig.~\ref{fig:M_L} the average mass of the light fragment in the outgoing channel of $^{40}$Ca+$^{238}$U at $E_{c.m.}=206$~MeV as a function of the initial angular momentum. 
Note that the mass of the fragments are determined as expectation values of one-body operators. 
Distribution of probabilities around these values could be extracted at the TDHF level thanks to particle-number projection techniques~\cite{sim10b}. However, the calculation of such distributions is going beyond the scope of this paper. 

We see in Fig.~\ref{fig:M_L} that all collisions with the tip lead to quasi-fission with partial mass equilibration. 
In particular, this orientation never leads to fusion, while the other orientation produces long contact time which may lead to fusion at $L\le10\hb$. 
In fact, collisions with the tip favour the production of a heavy fragment in the $^{208}$Pb region up to $L=70\hb$ which indicates that these QF reactions are strongly affected by shell effects. 
On the contrary, collisions with the side does not seem to produce an excess of $^{208}$Pb. 
This may be due to the fact that more compact configurations are reached with this orientation, favouring more mass symmetric exit channels for the most central collisions. 
In addition, QF is observed for collisions with the side up to $L=30\hb$ only.
For larger $L$, the overlap between the fragments is too small to allow the formation of a di-nuclear system, leading essentially to quasi-elastic reactions. 

\begin{figure}
\begin{center}
\includegraphics[width=8cm]{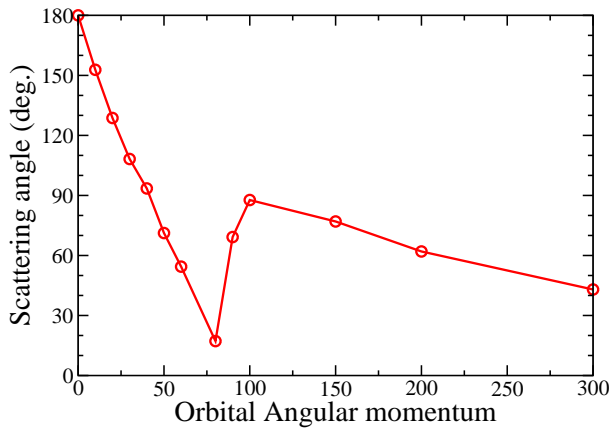}
\caption{Scattering angle as a function of angular momentum in collisions of a $^{40}$Ca with the tip of a $^{238}$U at $E_{c.m.}=206$~MeV.}
\label{fig:Theta_L}
\end{center}
\end{figure}

Usual experimental observables to investigate the quasi-fission process include, in addition to the fragment  mass distribution, the kinetic energy of the fragments \cite{itk11} and the scattering angle~\cite{tok85,tho08}.
Fig.~\ref{fig:Theta_L} shows the evolution of the scattering angle in $^{40}$Ca+$^{238}$U at $E_{c.m.}=206$~MeV for collisions with the tip of $^{238}$U as a function of angular momentum. 
For symmetry reasons, central collisions ($L=0\hb$) induce backward scattering at $180^\circ$, while
non-central collisions produce a rotation of the di-nuclear system.  
The higher the angular momentum, the higher the angular velocity. 
As a result, increasing $L$ induces quasi-fission with more forward angle down to $\sim17^\circ$ at $L=80\hb$ (see Fig.\ref{fig:densQF} left column for the associated density evolution). 
This indicates almost half a rotation before emission of the fragments at $L=80\hb$. 
For $L\ge100$, quasi-elastic scattering occurs with no orbiting, i.e., the nuclei follow a Coulomb trajectory.

The study of correlations between mass transfer and scattering angle, and comparisons with experimental data taken at ANU are ongoing~\cite{wak12}. 

\section{Conclusions and outlooks}

The quasi-fission process, which is responsible for the fusion hindrance in heavy system, strongly depends on the entrance channel properties. 
Measured mass angle distributions in several Ca,Ti+Pb,Hg systems  show narrower fragment mass distributions compatible with fusion when the colliding partners have strong shell effects and similar $N/Z$ ratio, such as in the $^{48}$Ca+$^{208}$Pb system. 
Microscopic mean-field calculations with the TDHF approach show that magic shells  also affect the outgoing channel by favouring the production of $^{208}$Pb-like fragments in the $^{40}$Ca+$^{238}$U reaction. 
It is also shown that the amount of mass transfer depends strongly on the orientation of the actinide. 

TDHF calculations of heavy-ion collisions provide observables, such as the mass of the fragments and the scattering angle, which can be directly compared to experimental data.  
However, a complete determination of mass-angle distributions requires  models which include beyond mean-field  fluctuations. For instance, extensions of TDHF calculations including fluctuations at the time-dependent RPA level are now available~\cite{sim11}. 
Alternatively, a stochastic mean-field approach could also be considered~\cite{was09b}.
These approaches should be applied to investigate quasi-fission in the near-future. 

\section*{Acknowledgements}

The authors are grateful to N. Rowley for motivating discussions. 
N. Lobanov and D. C. Weisser are thanked for intensive ion source development.
The calculations have been performed on the Centre de Calcul Recherche et Technologie of the Commissariat \`a l'\'Energie Atomique, France, and on the NCI National Facility in Canberra,
Australia, which is supported by the Australian Commonwealth Government.

Partial support from ARC Discovery grants DP0879679,  
DP110102858, and DP110102879 is acknowledged.

\bibliography{biblio}

\end{document}